\documentclass[aps,twocolumn,superscriptaddress,longbibliography,amsmath,amssymb,amsfonts,citeautoscript]{revtex4-2}
\usepackage{graphicx}
\usepackage{bm}
\usepackage{color}
\usepackage{epstopdf}
\usepackage{amsmath}
\usepackage{amssymb}
\usepackage{ulem}
\usepackage{verbatim} 
\usepackage[urlcolor=blue,colorlinks=true,citecolor=blue,linkcolor=blue,pdfstartview={FitH},bookmarks=false]{hyperref}

\newcommand{\e}{\varepsilon}

\newcommand{\be}{\begin{equation}}
\newcommand{\ee}{\end{equation}}
\newcommand{\bea}{\begin{eqnarray}}
\newcommand{\eea}{\end{eqnarray}}

\newcommand{\w}{\omega}
\newcommand{\s}{\sigma}

\newcommand{\up}{\uparrow}

\newcommand{\T}{\mathcal{T}}

\begin{document}


\title{Spin-dependent signatures of Majorana modes in thermoelectric transport through double quantum dots}

\author{Piotr Majek}
\email{pmajek@amu.edu.pl}
\affiliation{Institute of Spintronics and Quantum Information, Faculty of Physics and Astronomy, Adam Mickiewicz University, 61-614 Pozna{\'n}, Poland}

\author{Ireneusz Weymann}
\email{weymann@amu.edu.pl}
\affiliation{Institute of Spintronics and Quantum Information, Faculty of Physics and Astronomy, Adam Mickiewicz University, 61-614 Pozna{\'n}, Poland}

\date{\today}

\begin{abstract}

We present a comprehensive theoretical analysis of the spin-dependent thermoelectric properties of a double quantum dot system coupled to a topological superconducting nanowire and ferromagnetic leads. The study focuses on the behavior of the Seebeck coefficient and its spin-resolved counterparts, with calculations performed by means of the numerical renormalization group method. We investigate the low-temperature transport regime, where a complex interplay between the two-stage Kondo effect, the ferromagnet-induced exchange field, and the Majorana coupling occurs. We demonstrate that thermoelectric measurements can reveal unique signatures of the Majorana interaction that are challenging to isolate in conductance measurements alone. It is shown that the exchange field fundamentally alters the thermoelectric response, leading to a rich, non-monotonic temperature evolution of the thermopower, which is driven by a temperature-dependent competition between the spin channels. Furthermore, we have identified qualitatively different regimes of spin thermopower generation, controlled by the interplay between the Majorana-induced asymmetry and the spin polarization of the leads. Finally, by connecting the system's thermoelectric response to the underlying transport asymmetries quantified by the conductance spin polarization, we provide a consistent and unified physical picture, proposing thermoelectric transport as a sensitive probe for Majorana signatures.

\end{abstract}

\maketitle

\section{Introduction}
One of the leading directions in the modern condensed matter physics research is a search for exotic quasi-particles with non-trivial topological properties \cite{Aguado2017}. Majorana fermions, particles identical to their own antiparticles, are of particular interest as they can emerge in low-dimensional systems as zero-energy modes \cite{Majorana1937Apr, Kitaev2001}. The fundamental importance of these Majorana Zero Modes (MZMs) extends beyond academic curiosity. Their non-Abelian exchange statistics is anticipated to provide a foundation for topological qubits that are robust to decoherence, positioning them as a promising platform for fault-tolerant quantum computation \cite{Nayak2008Sep, MOORE1991362, Kitaev2003, Bonderson2008, aasen2025}.

Currently, a prominent platform for the realization of MZMs are semiconductor-superconductor heterostructures \cite{Lutchyn2010Aug, Oreg2010Oct}. In such systems, the appearance of Majorana fermions is supported by the zero bias peaks (ZBP), that are the signatures of the bound, ingap states fixed to the zero energy, present for a wide range of gate voltages and external magnetic fields \cite{Mourik2012May, Das2012Nov, Deng2012Dec}. Despite significant progress in the fabrication of such platforms, the unambiguous identification of MZMs remains challenging due to non-topological Andreev bound states, whose signatures are similar to MZMs \cite{Krogstrup2015, Chang2015, Kells2012Sep, Prada2012Nov, Pikulin2012Dec, Vuik2019, Pan2020}. To meet this challenge, rigorous verification schemes, such as the Topological Gap Protocol (TGP), have been proposed \cite{pikulin2021, Aghaee2023}. However, even this demanding protocol has faced criticism due to the similarity of signals from trivial states that can mimic their topological analogues, questioning the unambiguous interpretation of the results \cite{Hess2023, Garisto2025}.

In light of these challenges, a promising and more controlled alternative is the bottom-up approach, where multiple quantum dots are coupled through superconductors to realize a discrete Kitaev chain \cite{Sau2012, leijnse_ParityQubits_2012}. Recent experiments on double and triple quantum dot chains show that a long enough system could hold a pair of stable (poor man's) Majorana bound states \cite{dvir_realization_2023, ten_haaf_two-site_2024, bordin_enhanced_2025, ten_haaf_observation_2025, bordin2025probingmajoranalocalizationphasecontrolled}. Such systems can evolve into more complex \textit{tetrons}, where four MZMs form the basis for a (topological) qubit \cite{Karzig2017, Plugge_2017}. This approach has recently culminated in a demonstration of fermion parity measurement, enabling the construction of the \textit{Majorana 1}, a topological quantum chip where quantum dots play a crucial role as a measurement device \cite{aghaee_interferometric_2025, Garisto2025}.

Quantum dots systems with a side-coupled Majorana wires become a key platform to study the fundamental MZMs properties \cite{Deng2016Dec}. The leakage of a Majorana quasiparticle into the dot leads to unique transport signatures, such as fractional conductance values \cite{Vernek2014Apr}. Moreover, in the strong correlation regime, a complex competition arises between the MZM and the two-stage Kondo effect that develops in the double quantum dot system \cite{Lee2013Jun, weymann_majorana-kondo_2020, PustilnikPRL01, vojtaPRB02, cornagliaPRB05, Chung2008Jan, SasakiPRL09, Wojcik2015Apr}. As we have shown in our previous work, the use of ferromagnetic contacts allows for the study of spin-selective transport, revealing how the ferromagnet-induced exchange field competes with the Majorana coupling \cite{majek_spin-selective_2024, martinekPRL03}. We have demonstrated that even a small lead polarization can profoundly impact the transport characteristics, leading to an almost complete suppression of the conductance in one of the spin channels and allowing for the tuning of the current's spin polarization sign \cite{majek_spin-selective_2024}.

While investigations of the interplay between Majorana physics and Kondo correlations in these systems have predominantly focused on electrical conductance, thermoelectric transport provides a complementary and highly sensitive probe of the system's electronic structure \cite{Lee2013Jun, weymann_majorana-kondo_2020, Weymann2017Jan, Majek2022}. The thermopower (Seebeck coefficient) is, by its nature, exceptionally sensitive to asymmetries in the density of states around the Fermi energy, a property rooted in the Mott relation which links it to the energy derivative of the transmission function \cite{Jonson1980, CostiZlatic}. This sensitivity has been experimentally exploited to investigate the Kondo resonance in quantum dots \cite{Dutta2018Dec}. An analysis of the thermopower can therefore provide new, unique information on the nature of the complex competition between electronic correlations and Majorana quasiparticles.

In this paper, we present a comprehensive theoretical analysis of the thermoelectric properties of a double quantum dot system coupled to a Majorana wire and forming a tunnel junction with ferromagnetic leads. Extending our previous work on conductance \cite{majek_spin-selective_2024}, we extend the analysis to include the Seebeck coefficient and its spin-resolved counterparts. Using the numerical renormalization group method \cite{Wilson1975, Bulla2008}, we study how the competition between the two-stage Kondo effect, the exchange field, and the Majorana coupling manifests in the thermoelectric properties. We show that thermoelectric measurements can reveal characteristic signatures of the Majorana interaction that are difficult to isolate in conductance measurements alone, thus providing new insight into the physics of this complex hybrid system.

\section{Model and Method}

\begin{figure}
	\centering
	\includegraphics[width=0.7\linewidth]{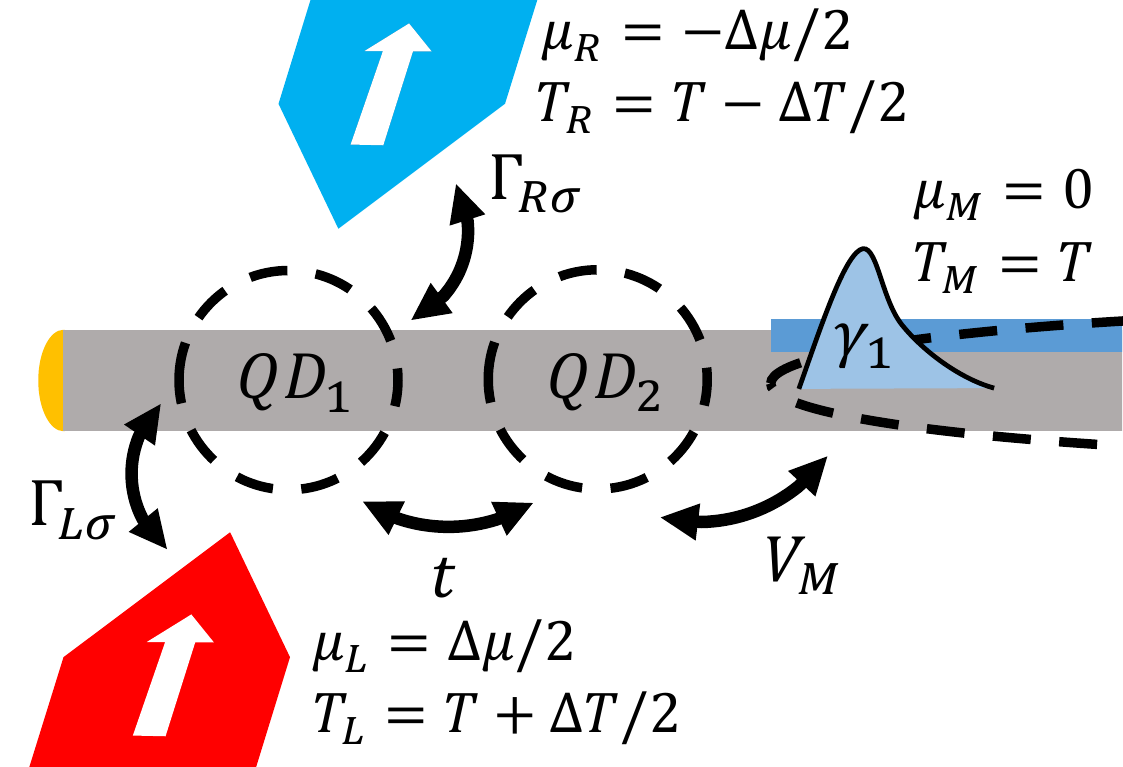}
	\caption{Schematic of the considered system. It consists of a double quantum dot forming a T-shaped geometry, where one of the quantum dots (QD$_1$) is directly coupled to external ferromagnetic leads, while the second quantum dot (QD$_2$) interacts with the topological superconducting nanowire hosting Majorana zero modes $\gamma_1$ and $\gamma_2$. In this work, we assume that the nanowire is long enough to neglect the Majorana mode $\gamma_2$ on the opposite end. There is a temperature gradient $\Delta T$ and chemical potential gradient $\Delta \mu$ applied between the ferromagnetic leads, while the topological superconductor is grounded and kept at temperature $T$.}
	\label{fig:Fig1}
\end{figure}

The considered system is illustrated in Fig. \ref{fig:Fig1}. It consists of a double quantum dot, which is attached to two ferromagnetic leads and side-coupled to a topological superconducting nanowire. More specifically, the first quantum dot is coupled to ferromagnetic leads, while the second dot is coupled to the Majorana mode emerging at the edge of the topological nanowire. To induce thermoelectric transport, the temperature and chemical potential gradients are applied symmetrically between the leads, while the nanowire is grounded. We focus on the linear response regime with respect to the applied potential and temperature gradients. Then, the thermoelectric coefficients can be related to the Onsager integrals \cite{barnard1972thermoelectricity}, which we calculate by means of the numerical renormalization group (NRG) procedure \cite{Wilson1975,Bulla2008}. This allows us to examine the system's behavior in a very accurate manner and, thus, shed light on the interplay of various correlations associated with the Kondo and Majorana physics, as well as with spin-resolved thermal transport and ferromagnetic-contact-induced exchange fields.

\subsection{Hamiltonian}
The total Hamiltonian of the system takes the following form:
\be
H = H_{\rm L} + H_{\rm T} + H_{\rm DD} + H_{\rm M}.
\ee
The first term, $H_{\rm L} = \sum_{r=L,R}\sum_{\mathbf{k}\sigma} \e_{r\mathbf{k}\sigma} c^\dag_{r\mathbf{k}\sigma} c_{r\mathbf{k}\sigma}$, describes the non-interacting electrons of the $r$-th electrode, where $r \in \{L, R\}$ indicates the left and right lead, respectively. The operator $c^\dag_{r\mathbf{k}\sigma}$ ($c_{r\mathbf{k}\sigma}$) creates (annihilates) an electron in the $r$-th electrode with spin $\sigma$, momentum $\bf{k}$, and the corresponding energy $\e_{r\mathbf{k}\sigma}$.
The tunneling of electrons between the ferromagnetic electrodes and the first quantum dot (QD$_1$) is described by the tunneling Hamiltonian
\be
H_{\rm T} = \sum_{r=L,R}\sum_{\mathbf{k}\sigma} v_{r \s} \left(d^\dag_{1\s}
c_{r\mathbf{k}\sigma} + c^\dag_{r\mathbf{k}\sigma} d_{1\s} \right),
\ee
where $v_{r \s}$ are the momentum-independent tunneling matrix elements, and $d^\dag_{1\s} (d_{1\s})$ is the creation (annihilation) operator for an electron with spin $\s$ on the first quantum dot. The coupling to the leads s a finite level broadening of the quantum dot, defined by $\Gamma_{r\s} = \pi \rho_{r \s} | v_{r \s}^2 |$, with $\rho_{r\s}$ being the spin-dependent density of states of lead $r$. The density of states is assumed to be flat in the window of width $2D$ around the Fermi level and its half-width is used here as the unit of energy ($D \equiv 1$).
Our discussion is focused on the spin-dependent phenomena, which are due to the presence of ferromagnetic leads. Moreover, we assume that the magnetic moments of the electrodes point in the same direction forming a parallel magnetic configuration. Then, the spin imbalance can be modeled through the coupling strength defined by, $\Gamma_{r \s} = (1 + \s p_r) \Gamma_r$, with $\Gamma_r = (\Gamma_r^\uparrow + \Gamma_r^\downarrow)/2$, and $p_r$ denoting the spin polarization of the lead $r$. In our calculations, we assume the system is left-right symmetric, simplifying $\Gamma_L = \Gamma_R \equiv \Gamma/2$ and $p_L = p_R \equiv p$.

The double quantum dot sub-system is modeled by the following Hamiltonian
\bea \label{Eq:HDD}
H_{\rm DD} &=& \sum_{j=1,2}\sum_{\s} \e_j d_{j\s}^\dag d_{j\s}
+ \sum_{j=1,2} U_j d_{j\uparrow}^\dag d_{j\uparrow} d_{j\downarrow}^\dag 
d_{j\downarrow}
\nonumber\\
&&+ \sum_\s t (d_{1\s}^\dag d_{2\s} +  d_{2\s}^\dag d_{1\s}),
\eea
where the operator $d_{j\s}^\dag$ ($d_{j\s}$) creates (annihilates) an electron with energy $\e_j$ on the $j$-th quantum dot (for $j = 1,2$). Here, $U_j$ describes the Coulomb interaction between electrons with the opposite spins on the $j$-th quantum dot, and $t$ models the tunneling of the spin-$\s$ electrons between the quantum dots.

Finally, the last term, $H_M$, describes effectively the topological superconducting nanowire and its coupling to the second quantum dot, where one of the Majorana quasiparticles, denoted by the operator $\gamma_1$, interacts with the QD$_2$ spin-down electrons with coupling strength $V_M$ \cite{Flensberg2010Nov,Liu2011Nov,Lee2013Jun}
\be \label{Eq:HM}
H_{\rm M} = \sqrt{2} V_M (d^\dag_{2\downarrow} \gamma_1 + \gamma_1 d_{2\downarrow}) + i \e_M \gamma_1 \gamma_2.
\ee
The second term in $H_M$, $i \e_M \gamma_1 \gamma_2$, introduces the second Majorana mode $\gamma_2$ and describes the overlap of their wavefunctions, denoted by $\e_M$. In our calculations, we consider the long nanowire case, where the overlap between the two Majorana modes can be neglected by setting $\e_M = 0$. This simplifies the Hamiltonian to $H_M =  \sqrt{2} V_M (d^\dag_{2\downarrow} \gamma_1 + \gamma_1 d_{2\downarrow})$. Finally, it is worth noting that Majorana operators satisfy the usual anticommutation relations, ${\{ \gamma_i, \gamma_j \} = \delta_{ij}}$, and can be represented in terms of an auxiliary fermionic operator $f$ as: $\gamma_1 = (f^\dagger + f)/\sqrt{2}$ and $\gamma_2 = i (f^\dagger - f)/\sqrt{2}$, which obeys the anticommutation relations, thus simplifying further calculations.

\subsection{Thermoelectric coefficients}
This paper focuses on the linear-response thermoelectric transport properties. Referring to Fig.\ref{fig:Fig1}, the temperature and voltage gradients, $\Delta T$ and $\Delta V$, are applied symmetrically between the left and right electrodes. The topological superconductor is assumed to be grounded ($ \mu_M = 0$) and kept at temperature $T_M = T$. The electric and heat currents can be then defined as follows: 
\be
\label{eq:current_matrix}
\begin{pmatrix}
	I \\
	I_{h}
\end{pmatrix}
= \sum_{\s}
\begin{pmatrix}
	e^2 L_{0\s}	&	-\frac{e}{T} L_{1\s} \\
	-e L_{1\s}	&	\frac{1}{T} L_{2\s}
\end{pmatrix}
\begin{pmatrix}
	\Delta V \\
	\Delta T
\end{pmatrix}.
\ee
The functions $L_{n\s}$ are the Onsager coefficients and can be expressed as:
\be
\label{eq:onsager_integral}
L_{n\s} = - \frac{1}{h} \int \w^n ~\frac{\partial f(\w)}{\partial \w} \T_\s (\w) d\w .
\ee
Here, $f(\w)$ denotes the Fermi-Dirac distribution function (for $\Delta \mu = \Delta T = 0$), and $\T_\s(\omega)$ is the spin-resolved transmission coefficient. The transmission coefficient can be related to the spectral function of the first quantum dot: $\T_\s (\w) = \pi \Gamma A_\s (\w)$, where $A_\s (\w) = -(1/\pi) \rm{Im} \mathcal{G}_\s^R (\w)$. Here, $\mathcal{G}_\s^R(\omega)$ is the Fourier transform of the retarded Green's function of the first quantum dot, which can be accurately calculated with the numerical renormalization group method.

In this paper, we focus on complementary transport coefficients: electric conductance $G$ and thermopower $S$ (also known as the Seebeck coefficient). Both quantities can be expressed as
\be
G_\s = \left( \frac{\partial I_\s}{\partial \Delta V} \right)_{\Delta T = 0} = e^2 L_{0\s},
\ee
with $G=\sum_\sigma G_\sigma$, and
\be
S_\s = \frac{1}{G_\s} \left( \frac{\partial I_\s}{\partial \Delta T}\right)_{\Delta V = 0} = - \frac{1}{e T} \frac{L_{1 \s}}{L_{0 \s}}.
\ee
The total thermopower can be then found from, $S = -(1/eT)(L_1/L_0)$, with $L_n = \sum_\s L_{n\s}$ \cite{barnard1972thermoelectricity}.

Furthermore, in the case of ferromagnetic leads with a long spin relaxation, there might be a spin voltage $\Delta V_{\rm{spin}}$ induced between the contacts, giving rise to finite spin thermopower \cite{Swirkowicz2009Nov,Weymann2013Aug,Wojcik2016Dec,Weymann2017Jan}.
The spin Seebeck effect $S_S$ can be found from
\be
S_S = - \frac{\Delta V_{\rm{spin}}}{\Delta T} = - \frac{1}{2 e T} \left(  \frac{L_{1 \up} - L_{1 \downarrow}}{L_{0 \up} + L_{0 \downarrow}} \right).
\label{eq:SS}
\ee

\section{Numerical results and discussion}

In the following section, we present a comprehensive analysis of the transport properties of the system. The section begins with the study of the linear conductance to establish the fundamental physical regimes governed by the interplay of the two-stage Kondo effect, Majorana coupling, and the ferromagnet-induced exchange field. Building upon this foundation, we then proceed to a detailed examination of the thermoelectric response, analyzing the charge thermopower and, subsequently, the spin thermopower. Finally, we summarize the underlying transport asymmetries by examining the conductance spin polarization, thereby providing a consistent physical picture. 

\subsection{Conductance}
\begin{figure}
	\centering
	\includegraphics[width=0.95\linewidth]{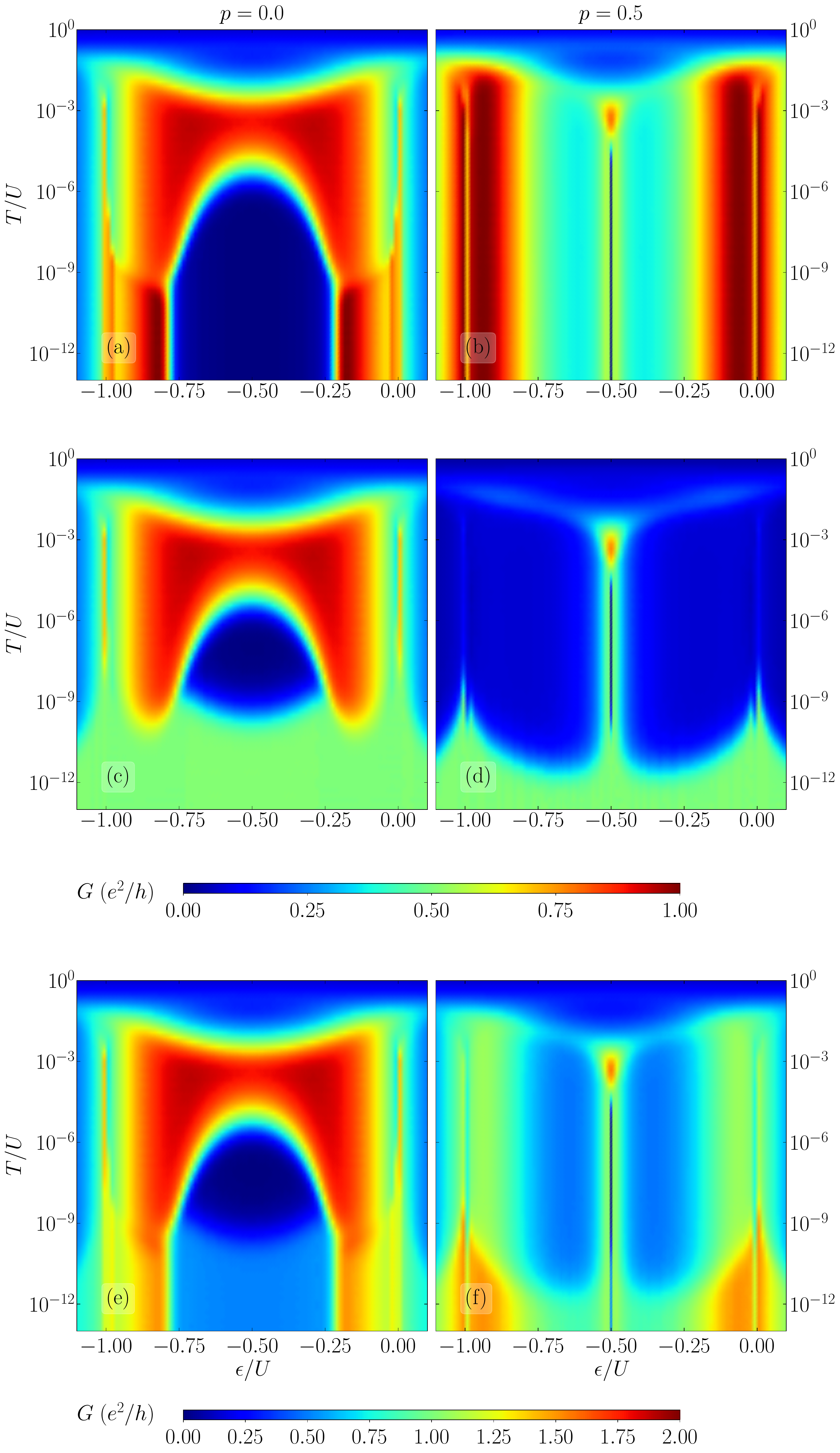}
	\caption{(a, b) Spin-up, (c, d) spin-down, and (e, f) total conductance of the system.
    The conductance is presented as a function of temperature $T$ and the energy levels of the quantum dots $\e_1 = \e_2 = \e$. Panels a, c, e in the left column correspond to non-magnetic leads ($p = 0$), whereas panels b, d, f in the right column represent ferromagnetic leads with $p = 0.5$.
		It is important to note that the conductance scale for individual spin contributions (panels a-d) ranges from 0 to $e^2/h$, while the total conductance (panels e, f) spans from 0 to $2e^2/h$, reflecting the sum of both spin channels.
		The calculations were performed with the following parameters: coupling to the Majorana nanowire $V_M = 10^{-5}U$, coupling to the leads $\Gamma = 0.1U$, hopping parameter $t = 0.02U$, and Coulomb correlation $U=0.2$ in units of the half-width of the flat density of states around the Fermi level, $D \equiv 1$.}
	\label{fig:Fig2}
\end{figure}

Here, to establish the basis for the discussion of spin-dependent thermoelectric properties, we examine the behavior of the linear-response conductance building upon previous work on the electronic properties of the system \cite{Majek2022}. The conductance as a function of the quantum dot energy level $\varepsilon$ and the temperature $T$ is illustrated in Fig. \ref{fig:Fig2}. This figure depicts the non-magnetic ($p=0$) and magnetic ($p=0.5$) cases in the left and right columns, respectively. To highlight the spin-resolved properties of the system, spin-up (top row) and spin-down (middle row) contributions are shown, with their combination as the total conductance presented in the bottom row. The energy levels are shifted simultaneously ($\varepsilon_1 = \varepsilon_2 = \varepsilon$), while the coupling to the Majorana nanowire and the hopping between the quantum dots are kept fixed, assuming: $t = 0.02U$ and $V_M = 10^{-5} U$. The leads are coupled to the first quantum dot, with the coupling strength $\Gamma = 0.1U$,
while $U=0.2$.

Upon examining Fig. \ref{fig:Fig2}, the analysis begins with the spin-up contribution to the conductance presented in the first row (panels a, b). Panel (a) illustrates the non-magnetic case ($p=0$), which serves as a reference. Here, characteristic energy scales of a strongly correlated double quantum dot system can be identified. Around the particle-hole symmetry point ($\varepsilon = -U/2$), at temperatures $T \approx \Gamma$, a Coulomb blockade regime is observed, characterized by small conductance peaks at Hubbard resonances, at $\e=0$ and $\e=-U$. In this regime, electronic transport is limited due to strong electrostatic repulsion between the electrons within the quantum dot. 

Lowering the temperature, the Kondo effect occurs, where a localized magnetic moment (unpaired electron spin in a quantum dot) is screened by conduction electrons from the leads. This screening results in a sharp resonance in the local density of states at the Fermi level, leading to an enhancement of conductance. This conductance enhancement develops around the Kondo temperature, $T\approx T_K$, which quantifies the energy scale of the Kondo correlations. 
The Kondo temperature for assumed parameters and for $\e=-U/2$, $t=0$ is $T_K \approx 0.0044\;U$. However, further decrease of temperature results in the suppression of the conductance for temperatures $T\lesssim T^*$, where $T^*$ is the second-stage Kondo temperature.

The influence of the Majorana nanowire becomes particularly relevant for the spin-down contribution depicted in panel (c). According to Eq.~(\ref{Eq:HM}),
the Majorana quasiparticle $\gamma_1$ is directly coupled to the spin-down electrons. This coupling significantly impacts the spin-down contribution to the conductance in the low-energy limit. Specifically, for temperatures $T \propto V_M^2$, the spin-down contribution to the conductance is observed to increase to $e^2/2h$ \cite{weymann_majorana-kondo_2020}, which is described by the energy scale $\Gamma_M$. In this regime, a competitive interplay between Majorana and Kondo physics becomes apparent. This competition leads to a dominant effect where even low values of the Majorana coupling destroy the second stage of the Kondo screening.

The impact of ferromagnetic leads ($p=0.5$) on the conductance is illustrated in the right column of Fig.~\ref{fig:Fig2}. A prominent observation across all spin contributions is the significant modification of the conductance features, primarily due to the introduction of spin polarization. 

Panel (b) reveals the spin-up contribution to the conductance. In this magnetic configuration, the two-stage Kondo effect, including its characteristic energy scales ($T_K$ and $T^*$), is significantly suppressed across the majority of the parameter space, except for the particle-hole symmetry point ($\varepsilon =-U/2$). At this point, the conductance peak is narrowed within the temperature range $10^{-2} \lesssim  T/U \lesssim 10^{-4}$. Away from the particle-hole symmetry point, the conductance peak, which was previously well-defined by the two-stage Kondo effect, now appears broadened across the shown temperature scale, particularly for the energy levels $-0.75 \lesssim \varepsilon /U \lesssim -0.25$.

In contrast, the spin-down contribution presented in panel (d) reveals that the introduction of spin polarization effectively suppresses the two-stage Kondo effect. Except for the particle-hole symmetry point, the conductance is significantly suppressed across most of the analyzed range of both temperature and quantum dot energy levels. Nevertheless, the influence of the Majorana coupling is observed, with its characteristic energy scale shifted towards considerably lower temperatures, from $T \approx 10^{-9}U$ to $T \approx 10^{-11}U$, thereby increasing the conductance to $G=e^2/2h$.

Panels (e) and (f) display the total conductance for both non-magnetic and magnetic cases. The introduced spin polarization of the leads significantly alters the overall conductance profile compared to the non-magnetic scenario. As a result, the two-stage Kondo effect, along with its characteristic energy scales, $T_K$ and $T^*$, vanishes. Nevertheless, the signature of the Majorana coupling (associated with $\Gamma_M$) remains present, although shifted toward lower temperatures. This indicates that for the given parameters, the competition between Kondo and Majorana physics is diminished. However, Majorana bound states manifest their presence even when the spin polarization is substantial.

\subsection{Thermopower}

\begin{figure}
	\centering
	\includegraphics[width=\linewidth]{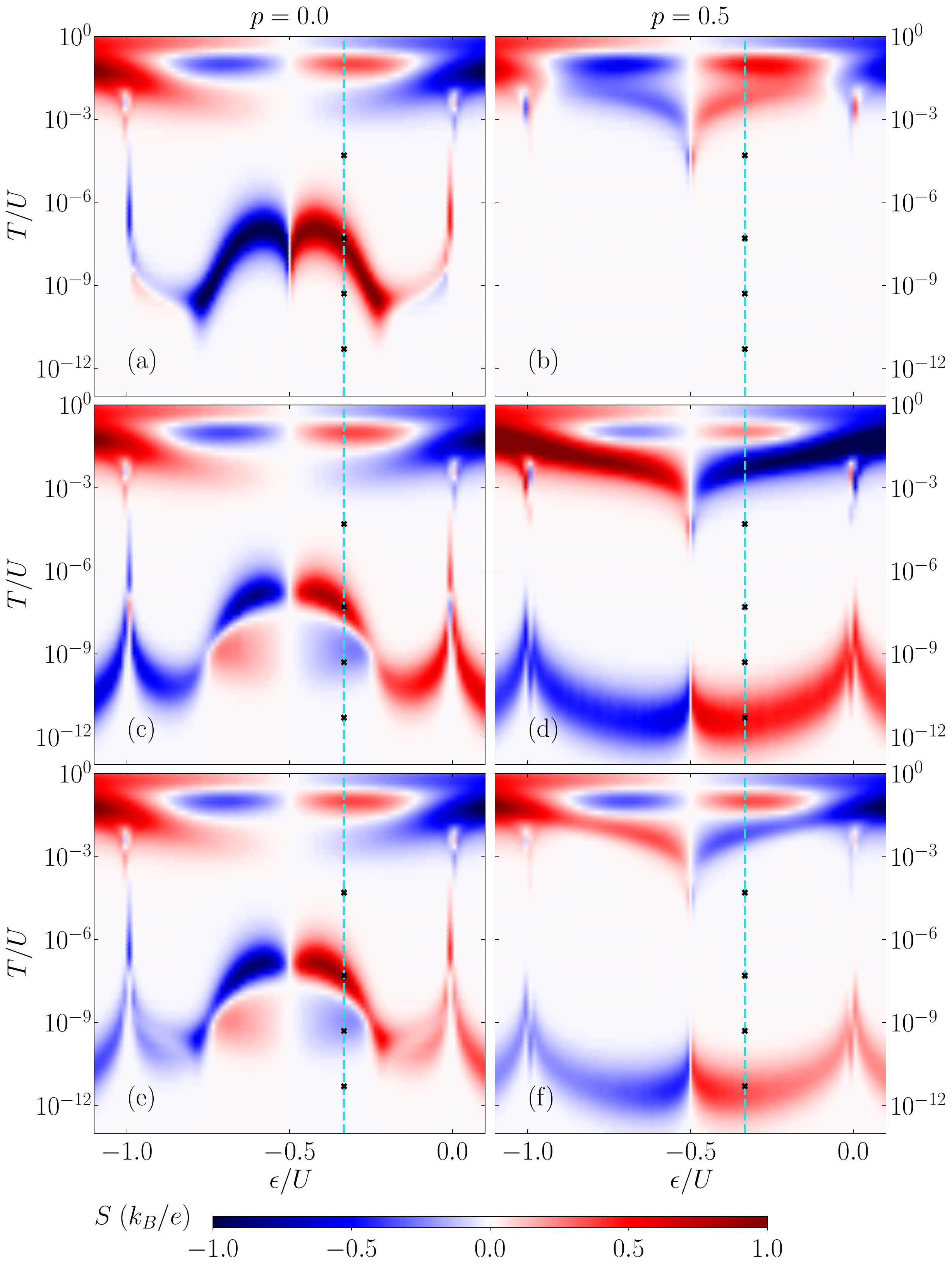}
	\caption{ (a, b) Spin-up, (c, d) spin-down, and (e, f) total thermopower of the system
    as a function of temperature and quantum dots' energy levels $\e_1 = \e_2 = \e$.
    The rest of the parameters is the same as in Fig.~\ref{fig:Fig2}.
    The dashed line indicates $\e = -U/3$ for reference to Figs.~\ref{fig:Fig4}, \ref{fig:Fig6}
    and \ref{fig:Fig7}.}
	\label{fig:Fig3}
\end{figure}

Following the detailed analysis of the conductance, attention is now turned to its thermoelectric properties, specifically the thermopower (the Seebeck coefficient). The calculated thermopower is presented in Fig.~\ref{fig:Fig3}, adopting a similar scheme to the conductance results
and utilizing the same set of parameters. Panels (a) and (b) show the spin-up thermopower for non-magnetic ($p=0$) and ferromagnetic ($p=0.5$) cases, respectively. Correspondingly, panels (c) and (d) show the spin-down thermopower, and panels (e) and (f) illustrate the total thermopower.

The first column, panels (a), (c), and (e), serves as a reference showing the thermopower induced by the temperature gradient applied to the non-magnetic leads. For More detailed analysis see \cite{Majek2022}.

For the efficiency of the present analysis, our attention is first
directed to the total thermopower for the non-magnetic case, illustrated in panel (e). For the temperature $T \approx \Gamma$, the thermopower exhibits peaks or dips of similar absolute magnitude depending on the energy relative to the particle-hole symmetry point. It is worth noting that for $\varepsilon = -U/2$ the thermopower vanishes, regardless of temperature, due to particle-hole symmetry.

Away from the particle-hole symmetry point, the thermopower sign change
as a function of temperature is observed. Such a sign change,
corresponding to the change of the carriers,
can be related to the conductance by the Mott formula:
\be
S_\s \approx - \frac{\pi^2}{3} \frac{k_B^2}{e} \frac{T}{\T_\s(0)} \frac{\partial \T_\s (\w)}{\partial \w}.
\label{eq:mottformula}
\ee
Here, $k_B$ denotes the Boltzmann constant, $T$ is the temperature, and $\T_\s (\w)$ refers to the transmission coefficient through the system for spin $\s$. 

In the Coulomb blockade regime, characterized by a small conductance peak for $T \sim 10^{-1}U$,
the corresponding feature is observed in the thermopower,
often manifesting as a sharp peak or dip with an associated sign change.
As evident from panels (a) and (c), the contribution from spin-up and spin-down components is of a similar magnitude. Such a pattern is also noted for the energy scales related to the two-stage Kondo effect, $T_K$ and $T^*$, where the thermopower sign change occurs and can be related to the conductance peak resulting from the screening of the spin on the first and the second quantum dot, respectively. Another sign change can be observed at the temperature scale associated with Majorana coupling ($T \sim \Gamma_M$). This thermopower peak (or dip) is also observed in the panel (c), which presents the spin-down thermopower, $S_\downarrow$.

The discussed pattern is observed for the double-dot energy levels within the range $-0.75 \lesssim \varepsilon/U \lesssim -0.25$. Outside this region, the spin-up thermopower ($S_\uparrow$) exhibits a sign change, whereas the spin-down contribution ($S_\downarrow$) shows a single thermopower peak (or dip, depending on the energy level). However, as far as the total thermopower is concerned,
the spin-up and spin-down contributions do not fully compensate each other.
This leads to the observation of a double-peak structure,
without any sign change, for energy ranges $-1 \lesssim \varepsilon/U \lesssim -0.75$
and $-0.25 \lesssim \varepsilon/U \lesssim 0$.

This double-peak structure becomes significantly modified when spin polarization is introduced. The presence of ferromagnetic leads induces an effective exchange field, which splits the quantum dot's energy levels \cite{martinekPRL03}. This field, whose magnitude depends on both the lead polarization $p$ and the gate voltage $\varepsilon$, competes directly with the
other energy scales in the system, such as the Kondo temperature and the Majorana-induced features, fundamentally altering the thermoelectric response. At the temperature scale of $\T \approx \Gamma$, a similar thermopower pattern to the non-magnetic case is observed.
The Coulomb blockade regime induces a peak (or dip), often accompanied by a sign change.

Notably, the major contribution is attributed to the spin-down thermopower [panel (d)], where a similar sign change is observed. Panel (b), however, reveals the double peak (dip) structure of the opposite sign compared to the spin-down contribution. This complex, non-monotonic evolution of the spin-resolved features with polarization can be linked to the exchange-field-induced splitting of the double dot's triplet states. As established in our previous work, a specific value of polarization can make the energy of the triplet component comparable to the singlet state, leading to an enhancement of Kondo-related processes and a significant modification of the low-energy spectrum \cite{majek_spin-selective_2024}, which is then reflected in the thermopower. This mechanism is rooted in the complex many-body physics of the T-shaped double quantum dot in the two-stage Kondo regime \cite{Wojcik2015Apr}.

The total thermopower vanishes across a wide temperature range, specifically $10^{-3} \gtrsim T/U \gtrsim 10^{-9}$. Below this range, for temperatures $T \lesssim 10^{-9}$, another peak (or dip) is observed. However, within the analyzed temperature range, no sign change occurs. 

For the given parameters, the impact originating from the coupling to the Majorana nanowire (which previously manifested as a sign change due to increased conductance for $T \approx \Gamma_M$ in the non-magnetic case) is not clearly observed. This leads to a stretching of the thermopower pattern, which was observed within the range around $\varepsilon / U \approx -1$ and $\varepsilon/U \approx 0$ for the non-magnetic case, to the full range of double-dot energies, excluding the particle-hole symmetry point ($\varepsilon = -U/2$), where the thermopower vanishes.

\begin{figure*}
	\centering
	\includegraphics[width=0.7\linewidth]{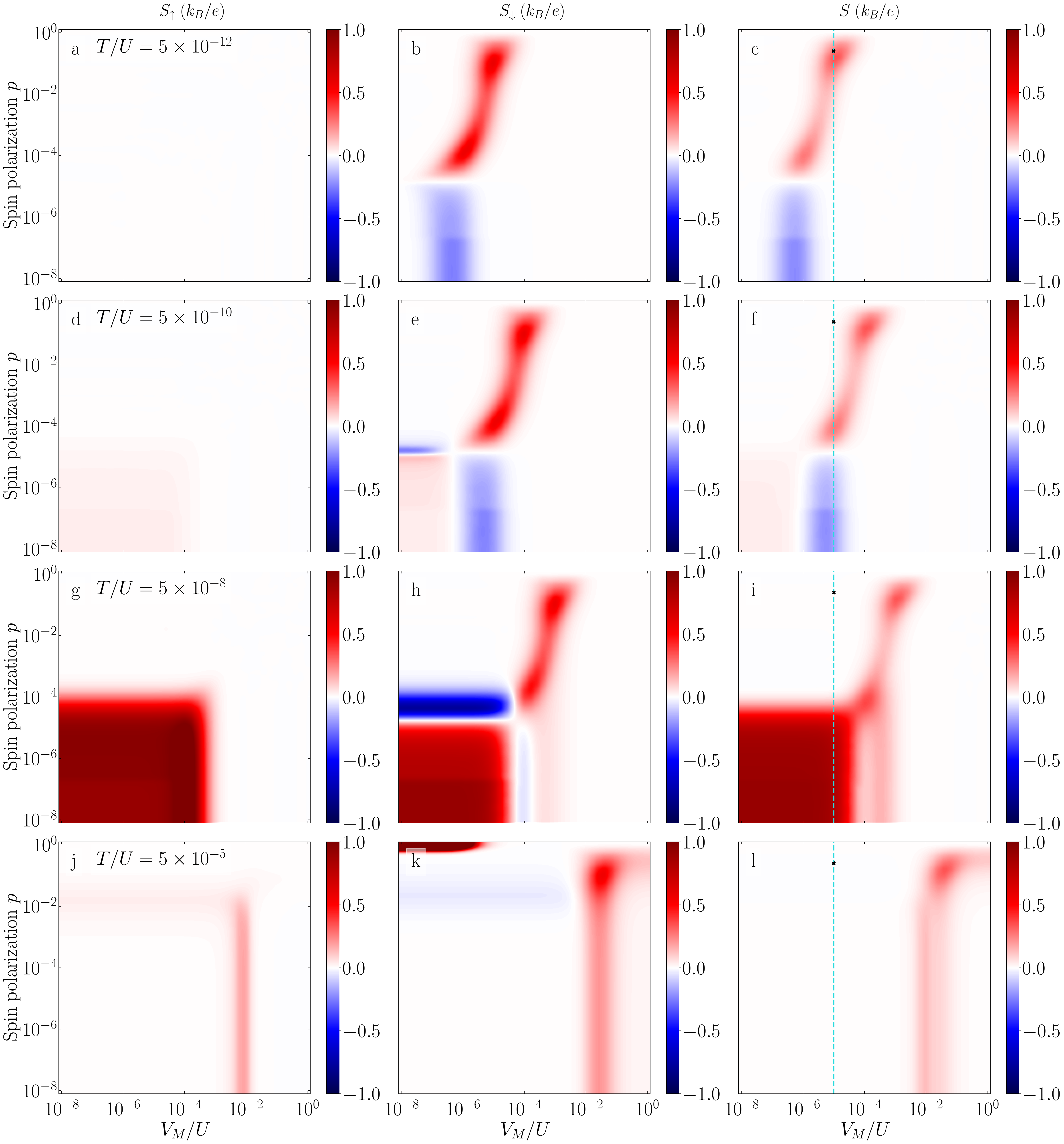}
	\caption{Thermopower as a function of spin polarization $p$ and temperature $T$. The first column (panels a, d, g, j) displays the spin-up thermopower, the second column (panels b, e, h, k) shows the spin-down contribution, and the third column (panels c, f, i, l) presents the total thermopower. All panels are calculated as a function of the spin polarization $p$ and the
    coupling strength to the Majorana mode $V_M/U$. Each row corresponds to a different temperature,
    as indicated. The dashed line in the last column indicates the Majorana coupling $V_M/U = 10^{-5}$ used in Fig.~\ref{fig:Fig3}; the black dot on this line highlights the $p \approx 0.5$. The rest of the parameters are the same as in Fig.~\ref{fig:Fig3}.}
	\label{fig:Fig4}
\end{figure*}

To further explore the tunability of these effects, particularly through experimentally relevant parameters, we now fix the dot energy level to $\e/U=-1/3$
and analyze the system's response in the $(p,V_M)$ parameter space.
This allows us to elucidate the combined influence of these two parameters
on the system's thermoelectric properties, as presented in Fig.~\ref{fig:Fig4}.

At the lowest temperature, $T/U = 5 \times 10^{-12}$ [panels (a-c)], the thermoelectric response is almost entirely dictated by the spin-down channel. As shown in panel (a), the spin-up contribution ($S_{\uparrow}$) is negligible. The total thermopower $S$ is therefore approximately equal to the spin-down component $S_{\downarrow}$. In this regime, we identify an optimal coupling range around $V_M/U \approx 10^{-6}$ that allows for efficient control over the thermopower sign via spin polarization. Here, the thermopower is negative for vanishing polarization ($p \to 0$) but becomes positive at $p \approx 10^{-5}$.

Increasing the temperature to $T/U = 5 \times 10^{-10}$ [panels (d-f)] reveals a more complex interplay between the spin channels. The most notable change is the emergence of a finite, positive spin-up contribution $S_{\uparrow}$ [panel (d)] at low $p$ and $V_M$. Although smaller in magnitude than the spin-down component, its influence is significant. It partially compensates the negative regions of $S_{\downarrow}$ [panel (e)], causing the area of negative total thermopower in panel (f) to be noticeably smaller. At this temperature, the system's response is highly tunable, with both $p$ and $V_M$ acting as parameters that can induce sign reversals.

A further increase in temperature to $T/U = 5 \times 10^{-8}$ [panels (g-i)]
leads to a qualitatively different thermoelectric landscape, characterized by the dominance of the spin-up channel. The spin-up contribution $S_{\uparrow}$ [panel (g)], previously weak, is now substantially amplified. This amplification is linked to the system's transport properties via the Mott formula, which in its simplified form relates the thermopower to the energy derivative of the transmission function $\mathcal{T}_{\sigma}(\omega)$, $S_{\sigma} \propto -(\partial \mathcal{T}_{\sigma}(\omega) / \partial \omega)|_{\omega=0}$. In this temperature regime, the system enters a second-stage screening process that strongly suppresses the spin-up conductance $G_{\uparrow}$ at the Fermi level. This suppression leads to a large energy derivative and, consequently, a large positive thermopower. This strong positive $S_{\uparrow}$ is large enough to overcome the negative regions of the spin-down contribution $S_{\downarrow}$ [panel (h)], resulting in a total thermopower that is exclusively positive. This phenomenon is found to be a robust feature, representing a fundamental change in the system's thermoelectric signature. Moreover, since the characteristic temperature of this second-stage Kondo effect, $T^*$, depends strongly on the interdot hopping $t$, the position of the thermopower maximum is also expected to be tunable via this parameter \cite{Wojcik2015Apr}.

Finally, at the highest analyzed temperature, $T/U = 5 \times 10^{-5}$ [panels (j-l)], the response is largely suppressed due to almost complete cancellation between the two spin channels. In the high-polarization regime, a faint, positive horizontal band in $S_{\uparrow}$ [panel (j)] is effectively counteracted by a more pronounced negative horizontal band in $S_{\downarrow}$ [panel (k)]. A robust thermoelectric signal survives only in a narrow vertical band for higher couplings
($V_M/U \approx 10^{-2}$), where both spin channels contribute positively, leading to a constructive reinforcement.

In summary, the presented figure illustrates a rich, non-monotonic temperature evolution of the thermopower, driven by a complex, temperature-dependent competition between the two spin channels.

\subsection{Spin thermopower}
\begin{figure}
	\centering
	\includegraphics[width=\linewidth]{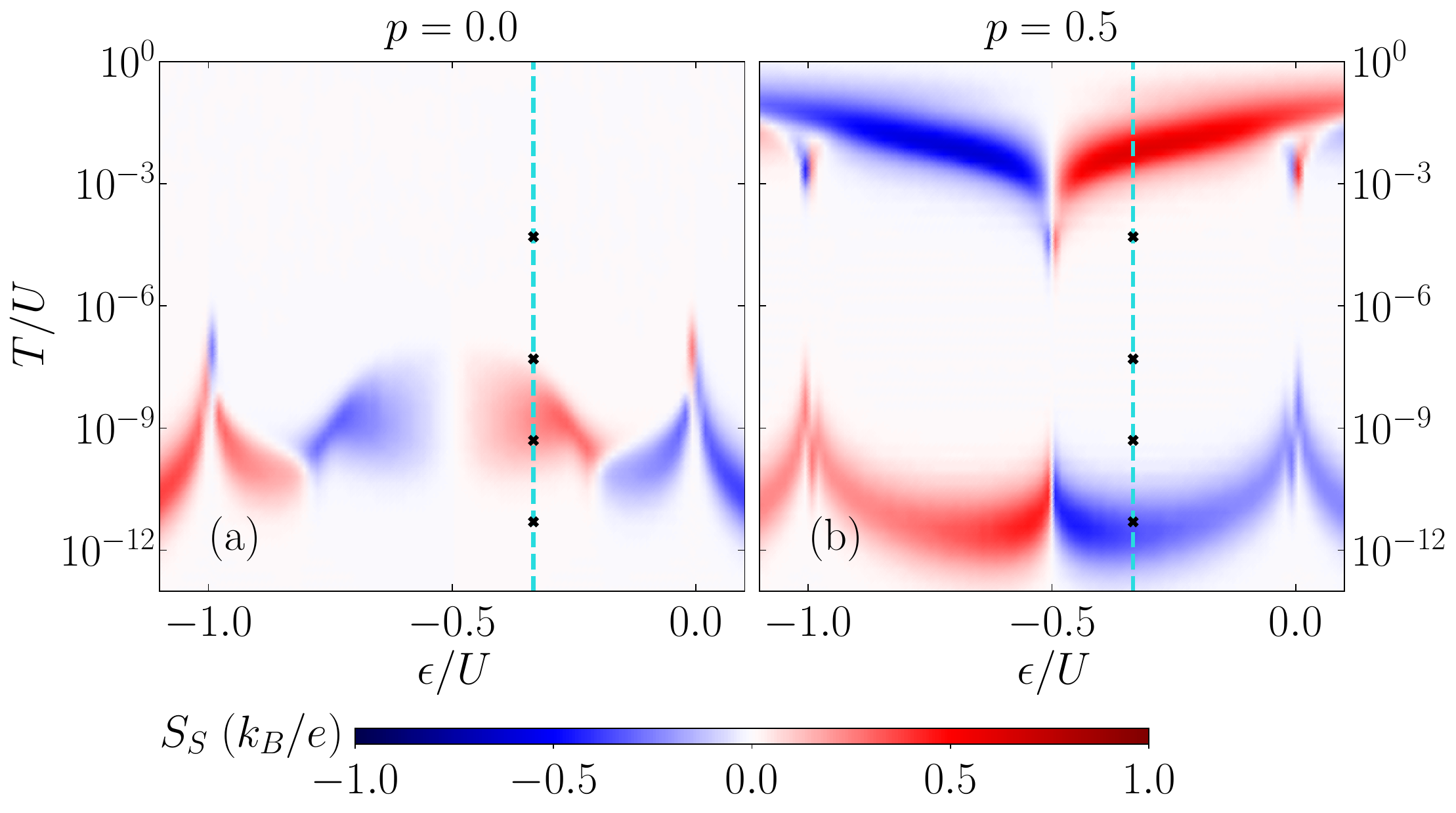}
	\caption{Spin thermopower as a function of temperature and energy level. The temperature and energy level dependence of the total spin thermopower calculated for the first quantum dot for the non-magnetic case ($p = 0$) and the magnetic case ($p = 0.5$). }
	\label{fig:Fig5}
\end{figure}
To quantify the efficiency of spin-polarized thermoelectric transport, we now investigate the spin thermopower, defined by Eq.~(\ref{eq:SS}). Its dependence on the temperature and dot energy level is presented in Fig.~\ref{fig:Fig5}, which contrasts the non-magnetic case ($p=0$, panel [a]) with a representative spin-polarized case ($p=0.5$, panel [b]).

In the absence of spin polarization [Fig.~\ref{fig:Fig5}(a)], the case described in more detail in Ref.~\cite{Majek2022}, a finite spin thermopower emerges only within
a specific low-temperature window, approximately $10^{-12} \lesssim T/U \lesssim 10^{-6}$.
The boundaries of this temperature window are determined by the competition
between the second-stage Kondo screening, which suppresses transport at higher temperatures,
and the Majorana coupling, which restores it at lower temperatures.
As expected, the spin thermopower is perfectly antisymmetric
with respect to the particle-hole symmetry point at $\varepsilon = -U/2$.

Introducing the spin polarization [Fig.~\ref{fig:Fig5}(b)] fundamentally alters this landscape. At low temperatures ($T/U \lesssim 10^{-6}$), while the perfect antisymmetry is retained, the region of vanishing thermopower around the particle-hole symmetry point becomes substantially narrower compared to the non-magnetic case. Most strikingly, the spin polarization induces a new regime of finite spin thermopower at high temperatures ($T/U \gtrsim 10^{-5}$), separated from the low-temperature features by a gap around $T/U \approx 10^{-6}$ where $S_s$ vanishes.
This high-temperature signal exhibits a complex structure with sharp features and sign changes near $\varepsilon/U \approx 0$ and $\varepsilon/U \approx -1$.

The physical origins of this evolution are rooted in the temperature-dependent interplay between the Majorana-induced asymmetry and the spin polarization of the leads. At low temperatures ($T/U \lesssim 10^{-6}$), the behavior is dominated by the hybridization with the Majorana modes, which breaks the symmetry between the spin channels and induces a finite $S_S$ even for $p=0$. As the temperature increases, this effect is suppressed by thermal fluctuations, and the role of the lead polarization becomes crucial. In the spin-polarized case ($p = 0.5$), the leads themselves introduce a strong, temperature-robust asymmetry. This results in the emergence of a large high-temperature spin thermopower, driven by the opposing contributions of the spin-up and spin-down channels, a feature entirely absent in the non-polarized system.

\begin{figure*}
	\centering
	\includegraphics[width=1\linewidth]{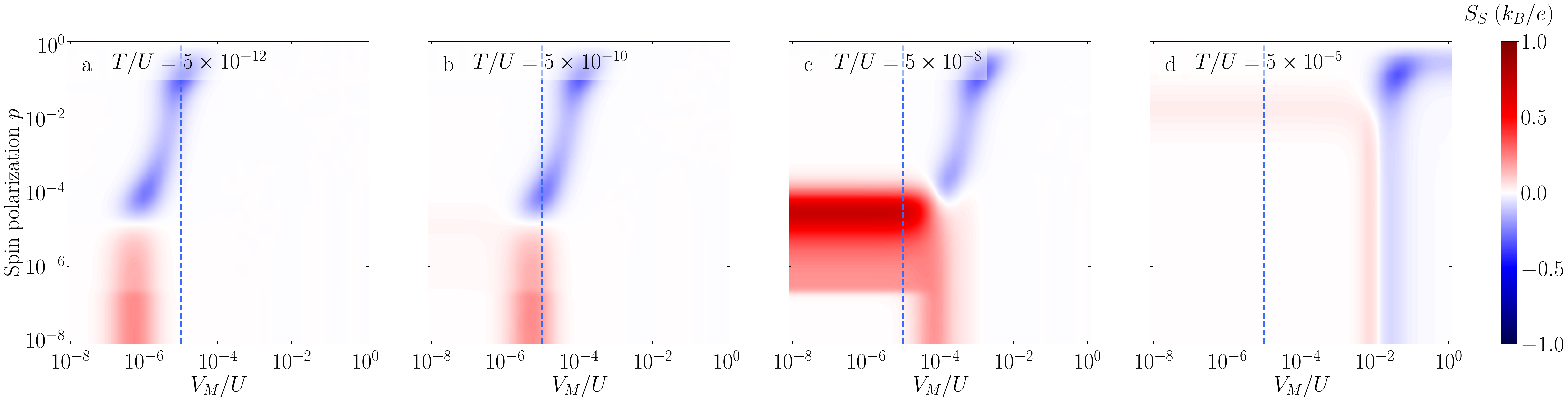}
	\caption{The spin thermopower as a function of the coupling to the topological
    nanowire $V_M/U$ and spin polarization of the leads,
    calculated for different temperatures. The temperatures and the rest
    of the parameters are indicated in Fig.~\ref{fig:Fig3}. }
	\label{fig:Fig6}
\end{figure*}

This analysis reveals distinct low- and high-temperature regimes for spin thermopower generation. To understand how these regimes are controlled by the Majorana coupling and lead polarization, in Fig.~\ref{fig:Fig6} we examine the $S_S$ landscape in the $(p,V_M)$ plane at representative temperatures. The underlying spin-resolved contributions are the same as those analyzed for the charge thermopower in Fig.~\ref{fig:Fig4}.

At the lowest temperatures, $T/U = 5 \times 10^{-12}$ [panel (a)] and $T/U = 5 \times 10^{-10}$ [panel (b)], the spin thermopower landscape closely mirrors that of the charge thermopower, but with an inverted sign. This observation can be understood by examining the individual spin channels. As shown in the corresponding panels of Fig.~\ref{fig:Fig3} [panels (b) and (d)], the spin-up contribution to the thermopower is negligible ($S_{\uparrow} \approx 0$) in this temperature regime. The total thermopower [Fig.~\ref{fig:Fig3}(f)] is thus dominated by the spin-down channel, so that $S \approx S_{\downarrow}$. Under these conditions, the spin thermopower reduces to $S_s \approx S_{\uparrow} - S_{\downarrow} \approx -S_{\downarrow}$, which simplifies to the observed relation $S_s \approx -S$. The signal is characterized by a positive region at low polarizations ($p \lesssim 10^{-5}$) and a negative region at higher polarizations, with both features being prominent for $V_M/U \approx 10^{-6}$. Increasing the temperature from panel (a) to (b) causes a slight shift of these features toward higher values of $V_M$.

A qualitative change occurs at the intermediate temperature of $T/U = 5 \times 10^{-8}$ [panel (c)], where the simple relation to the charge thermopower breaks down. A large region of strong positive spin thermopower now dominates the low-polarization regime ($p \lesssim 10^{-4}$). The sign-change boundary becomes more complex, showing a strong dependence on both $p$ and $V_M$. For instance, along the line of fixed coupling $V_M/U = 10^{-5}$ (dashed line), $S_S$ exhibits a pronounced peak structure, a feature not apparent in simpler cross-sections.

Finally, at the highest analyzed temperature, $T/U = 5 \times 10^{-5}$ [panel (d)], the spin thermopower pattern undergoes a further qualitative change. The signal is now suppressed for Majorana couplings below $V_M/U \approx 10^{-3}$ across the entire polarization range and becomes significant only for stronger couplings. The most striking feature is the nature of the sign reversal, which now occurs along a near-vertical line at a fixed coupling of $V_M/U \approx 10^{-2}$ that spans all values of $p$. A positive signal is observed for $10^{-3} \lesssim V_M/U < 10^{-2}$, while a negative signal appears in a band centered closer to $V_M/U \approx 2 \times 10^{-2}$ before being suppressed at even stronger couplings.

Taken together, Fig.~\ref{fig:Fig6} demonstrates how temperature transforms the spin thermopower from a simple inverted reflection of the charge thermopower at low temperatures into a signal governed by complex, distinct mechanisms of compensation and reinforcement at high temperatures.

\subsection{Conductance spin polarization}

\begin{figure*}[ht]
    \centering
    \includegraphics[width=1\linewidth]{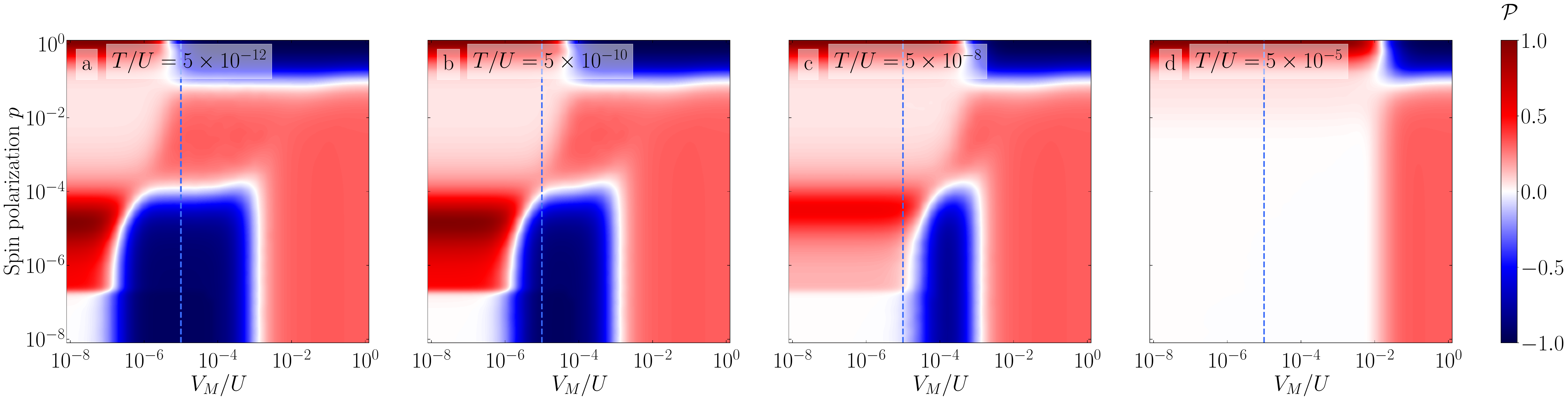}
    \caption{The conductance spin polarization calculated
    as a function of the lead spin polarization $p$
    and the coupling to Majorana wire $V_M$ at different temperatures,
    assuming $\varepsilon_1 = \varepsilon_2 = -U/3$.}
    \label{fig:Fig7}
\end{figure*}

Finally, it is insightful to connect these thermoelectric landscapes to the underlying spin transport asymmetries, which are quantified by the conductance spin polarization, $\mathcal{P} = (G_{\uparrow} - G_{\downarrow})/(G_{\uparrow} + G_{\downarrow})$. Figure \ref{fig:Fig7} displays the temperature evolution of $\mathcal{P}$ in the $(p, V_M)$ plane, showing the boundaries between regions of spin-up (red) and spin-down (blue) transport dominance. A key aspect of this figure is the evolution of the spin-down dominated region ($\mathcal{P}<0$) at low $p$ and low $V_M$. As temperature increases, its left boundary shifts towards higher $V_M$, while the right boundary remains anchored around $V_M/U \approx 10^{-3}$, thus narrowing this region. At $T/U = 5 \times 10^{-8}$ [panel (c)], it is reduced to a small area concentrated around $V_M/U \approx 10^{-4}$. It is important to note, however, that a separate region of spin-down dominance persists at high values of both $p$ and $V_M$.

This temperature-driven evolution of the spin compensation boundary is a primary driver of the changes observed in the thermoelectric response (Figs. \ref{fig:Fig5} and \ref{fig:Fig6}). The most significant thermoelectric features are consistently anchored to the $\mathcal{P}=0$ contour that separates the competing low-energy regimes. The gradual disappearance of the low-$p$, low-$V_M$ spin-down dominated region [Fig. \ref{fig:Fig7}(d)] is thus consistent with the previously discussed physical mechanism, where the spin-up channel's dominance is induced by temperature in this specific parameter range. This provides a consistent and unified physical picture, directly linking the features in electrical conductance to the system's thermoelectric signatures.

\section{Summary}
In this paper, we have presented a comprehensive theoretical analysis of the spin-dependent thermoelectric properties of a double quantum dot system coupled to a topological superconducting nanowire and ferromagnetic leads. Extending our previous work on the system's conductance, the study focuses on the behavior of the Seebeck coefficient and its spin-resolved counterparts. The calculations were performed by means of the numerical renormalization group method, which allowed for a reliable investigation of the low-temperature transport regime, where a complex interplay between the two-stage Kondo effect, the ferromagnet-induced exchange field, and the Majorana coupling occurs.
We have demonstrated that thermoelectric measurements can reveal unique signatures of the Majorana interaction that are challenging to isolate in conductance measurements alone. Our study shows that the exchange field induced by the ferromagnetic leads fundamentally alters the thermoelectric response, competing with other energy scales and modifying the low-energy spectrum. This results in a rich, non-monotonic temperature evolution of the thermopower, driven by a complex, temperature-dependent competition between the spin channels.
Furthermore, we have identified distinct temperature regimes for spin thermopower generation, which are controlled by the interplay between the Majorana-induced asymmetry and the spin polarization of the leads. Finally, by connecting these thermoelectric landscapes to the underlying transport asymmetries quantified by the conductance spin polarization, we have provided a consistent and unified physical picture of the system's behavior. Our findings offer new insights into the physics of this complex hybrid system and propose thermoelectric transport as a sensitive probe for Majorana signatures.

\section{Acknowledgments}

This work was supported by the National Science Centre in Poland through the Project No. 2021/41/N/ST3/01885.


%
	
\end{document}